\documentclass[12pt,preprint]{aastex}

\usepackage{epsfig}
\begin{document}
\title{Measuring the Abundance of sub-kilometer sized Kuiper Belt Objects
  using Stellar Occultations} 
 
 \shortauthors{Schlichting et al.}
  
\shorttitle{Sub-km sized Kuiper Belt Objects} \author{Hilke E. Schlichting\altaffilmark{1,2,3}, Eran O. Ofek\altaffilmark{4}, Re'em Sari\altaffilmark{5}, Edmund P. Nelan\altaffilmark{6}, Avishay Gal-Yam\altaffilmark{4}, 
Michael Wenz\altaffilmark{7}, Philip Muirhead\altaffilmark{2}, Nikta Javanfar\altaffilmark{8}, Mario Livio\altaffilmark{6}}\altaffiltext{1} {UCLA, Department of Earth and Space Science, 595 Charles E.  Young Drive East, Los Angeles, CA 90095}\altaffiltext{2} {California Institute of Technology, MC 130-33, Pasadena, CA 91125}\altaffiltext{3} {Hubble Fellow}\altaffiltext{4}{Faculty of Physics, Weizmann Institute of Science, POB 26, Rehovot 76100, Israel}\altaffiltext{5}{Racah Institute of Physics, Hebrew University, Jerusalem 91904, Israel}\altaffiltext{6}{Space Telescope Science Institute, 3700 San Martin Drive, Baltimore, MD 21218}
\altaffiltext{7}{Goddard Space Flight Center, 8800 Greenbelt Road, Greenbelt, MD 20771}
\altaffiltext{8}{Queen's University,
99 University Avenue, Kingston, Ontario K7L 3N6, Canada}
 \email{hilke@ucla.edu}

\begin{abstract} 
We present here the analysis of about 19,500 new star hours of low ecliptic latitude observations ($|b| \leq 20^{\rm{\circ}}$) obtained by the Hubble Space Telescope's Fine Guidance Sensors over a time span of more than nine years; which is an addition to the $\sim 12,000$ star hours previously analyzed by \citet{SO09}. 
Our search for stellar occultations by small Kuiper belt objects (KBOs) yielded one new candidate event corresponding to a body with a $530\pm{70}\rm{m}$ radius at a distance of about 40~AU. Using bootstrap simulations, we estimate a probability of $\approx 5\%$, that this event is due to random statistical fluctuations within the new data set. Combining this new event  with the single KBO occultation reported by \citet{SO09} we arrive at the following results: 1) The ecliptic latitudes of $6.6^{\circ}$ and $14.4^{\circ}$ of the two events are consistent with the observed inclination distribution of larger, 100\,km-sized KBOs. 2) Assuming that small, sub-km sized KBOs have the same ecliptic latitude distribution as their larger counterparts, we find an ecliptic surface density of KBOs with radii larger than 250~m of $N(r>250\,m) = 1.1^{+1.5}_{-0.7} \times 10^7\,\rm{deg^{-2}}$; if sub-km sized KBOs have instead a uniform ecliptic latitude distribution for $-20^{\circ} < b< 20^{\circ}$ then
$N(r>250\,m) = 4.4^{+5.8}_{-2.8} \times 10^6\,\rm{deg^{-2}}$. This is the best measurement of the surface density of sub-km sized KBOs to date. 3) Assuming the KBO size distribution can be well described by a single power law given by $N(>r) \propto r^{1-q}$, where $N(>r)$ is the number of KBOs with radii greater than $r$, and $q$ is the power law index, we find $q=3.8\pm{0.2}$ and $q=3.6\pm{0.2}$ for a KBO ecliptic latitude distribution that follows the observed distribution for larger, 100-km sized KBOs and a uniform KBO ecliptic latitude distribution for  $-20^{\circ} < b< 20^{\circ}$, respectively. 4) Regardless of the exact power law, our results suggest that small KBOs are numerous enough to satisfy the required supply rate for the Jupiter family comets. 5) We can rule out a  single power law below the break  with $q>4.0$ at 2$\sigma$, confirming a strong deficit of sub-km sized KBOs compared to a population extrapolated from objects with $r> 45~\rm{km}$. This suggests that small KBOs are undergoing collisional erosion and that the Kuiper belt is a true analogue to the dust producing debris disks observed around other stars.
\end{abstract}


\section{INTRODUCTION}
The Kuiper belt consists of a disk of icy objects and is located at the outskirts of our planetary system just beyond the orbit of Neptune. Since the discovery of the first Kuiper belt object (KBO) in 1992 \citep{JL92}
well over 1000 objects have been detected. Dedicated surveys have been revealing an
intricate dynamical structure of the Kuiper belt \citep[e.g.,][]{PKG11}, have been investigating the multiplicity of KBOs \citep[e.g.,][]{N08}, and have been measuring the Kuiper belt size-distribution \citep[e.g.,][]{BTA04}. In the Kuiper belt,
planet formation never reached completion because runaway growth of the
planetary embryos was interrupted, most likely, by an increase in velocity dispersion of the
planetesimals. The size distribution of larger KBOs provides a snap-shot of planet formation that was erased elsewhere in the solar system, where planet formation went all the way to completion.

The cumulative size distribution of KBOs larger than about $45~\rm{km}$ in radius is well
described by a single power-law given by
\begin{equation}\label{e1}
N(>r) \propto r^{1-q}
\end{equation} 
where $N(>r)$ is the number of objects with radii greater than $r$, and $q$ is
the power-law index. Observations find that the power-law index for large KBOs
is $q \sim 4.5$ \citep{FH108,FK108}. Since this size distribution is a relic of the
accretion history in the Kuiper belt, it provides valuable insights
into the formation of large KBOs and the planet formation process itself \citep[e.g.][]{S96,DF97,K02}. Coagulation
models of planet formation reproduce the observed size distribution of large
KBOs well \citep[e.g.,][]{KL99,K02,SS11}.

Observations suggest that there is a break in the power-law size distribution
at smaller KBO sizes \citep[e.g.][]{BTA04,FH108,FK108,SO09,FH10}. The break in
the size distribution is generally attributed to collisions that break-up
small KBOs (i.e.,$r \lesssim 45~\rm{km}$) and subsequently modify their size
distribution \citep[e.g.][]{D69,KB04,PS05}. If this is so, then the location
of the break (i.e., the break radius) constrains the time period over which
destructive collisions have been occurring in the Kuiper belt.  This is
because with time, collisions move the break radius to ever larger
sizes. Furthermore, the power-law index below the break radius constrains the
material properties of KBOs. Small KBOs that are in collisional
equilibrium and that are held together predominantly by material strength are expected to have
a size distribution with power-law index $q = 3.5$, which is the so-called
Dohnanyi spectrum \citep{D69}. The size distribution will be shallower with a
power-law index of $q \approx 3$ if small KBOs are held together predominantly by
gravity (i.e., they are effectively rubble piles) \citep{PS05}. The break
radius and the power-law index below the break, therefore, constrain the
collisional history of the Kuiper belt and reveal the material properties of
small KBOs, respectively.

Our interest in studying the abundance and size distribution of small ($r<10\rm{km}$) KBOs is threefold. First of all, we want to determine their overall abundance to test if they are numerous enough to supply the Jupiter Family comets. Second, we want to measure their size distribution in order to constrain collisional evolution in the Kuiper belt and establish it as a true analog of the dust producing debris disk observed around other stars.
And third, we would like to constrain the small KBO size distribution well enough to infer their material properties as discussed above.

Unfortunately, not much is known about the KBO size distribution significantly below the break,
because objects with radii less than $\sim 10\rm{km}$ are too faint to be detected in reflected light. These objects can, however, be detected indirectly by stellar occultations
\citep{B76,D92,AAC92}. A small KBO crossing the line of sight of a star will partially obscure
the stellar light, which, under suitable circumstances, can be detected in the star's light curve. 
Both optical \citep[e.g.][]{RDD06,BKW08,BP09,SO09,BZ10,WP10} and X-ray \citep{CKL06,CCC11} occultation
searches have been conducted to probe the population in the Kuiper belt. \citet{CKL06} searched for occultation signatures in archival RXTE X-ray data of Scorpius-X1 and reported a surprisingly high event
rate. However, \citet{JLM08} showed that most of the dips in the Scorpius-X1
light curves are artificial effects caused by the RXTE photomultiplier. Only 12 of the original 58 events were not ruled out as artifacts; however \citet{BKW08} point out that the duration and/or depth of most of the
12 remaining events are inconsistent with the diffraction signature of
occultation events. \citet{RDD06}, \citet{BKW08}, and \citet{WP10} all conducted occultation
surveys in the optical regime. None of these surveys reported any detections of
objects in the Kuiper belt, which is not surprising given the low expected
event rate of these surveys. Ground-based observations may suffer from a high
rate of false positives due to atmospheric scintillation. The only
ground-based system that is currently able to screen events caused by
atmospheric scintillation and other kinds of interference is the Taiwanese American Occultation Survey (TAOS), which consists of four telescopes that observe the same position simultaneously \citep[e.g.,][]{Aet03}. TAOS collected over 500,000 star hours with 4 or 5~Hz sampling frequency and reported no detections so far \citep{BZ10}. Recently \citet{DLR12} analyzed about 140,000 star hours sampled at 1~HZ by the COROT satellite and reported up to 15 possible KBO occultation events. The first measurement of the surface density of sub-km sized KBOs has been reported by \citet{SO09}, who analyzed 12,000 star hours of archival HST data with 40~Hz sampling frequency. \citet{SO09} reported one detection and found an ecliptic surface density of KBOs with radii larger than 250~m of $2.1^{+4.8}_{-1.7} \times 10^7~ \rm{deg^{-2}}$.

In this paper we present the analysis of new data obtained
by the Fine Guidance Sensors (FGSs) on board the Hubble Space Telescope
(HST) over a time span of more than nine years. We searched this data set for stellar occultations by small KBOs with the aim to measure their abundance, to constrain their size distribution and
to address the question of whether they are ubiquitous enough to satisfy the
supply needed to explain the observed frequency of Jupiter family comets.

This paper is structured as follows. We describe our HST-FGS survey in section
\ref{s22}. Details about our data analysis and detection efficiency follow in
sections \ref{s33} and \ref{s44}, respectively.  In section \ref{s55} we
present the results from our survey. We calculate the abundance of sub-km sized KBOs, compare our findings with other Kuiper belt surveys and conclude in section \ref{s66}.

\section{THE HST-FGS OCCULTATION SURVEY}\label{s22}
The HST Fine Guidance Sensors (FGSs) are an integral part of the HST pointing
control system, ensuring a pointing stability of HST at the milli-arc-second
level over exposure times of tens of minutes. Each FGS is a dual-axis white
light shearing interferometer and the photon count of each FGS is recoded by 4
PMTs (e.g. Fine Guidance Sensor Instrument Handbook: http://www.stsci.edu/hst/fgs/documents/instrumenthandbook/fgs\_ihb.pdf). In addition to ensuring the pointing stability of HST, the FGSs have been used
as science instruments. For example, the FGS was used as an astrometric instrument to measure the precise distances to eight galactic Cepheids \citep{BMF07}, as an interferometer to resolve binary O-stars in the Carina Nebula \citep{NWW04}, and as a high speed photometer to study transit light curves of the planet orbiting HD and asteroseismology of the star itself \citep{GMN11}.

For over a decade, the three FGSs have been collecting a large number of
photometric measurements of the guide stars with 40 Hz time resolution. In this paper we
search more than nine years of this archival data set for photometric signatures consistent with the
occultation of the guide star by a small object in the Kuiper belt. The 40 Hz
sampling frequency, the large number of star hours of FGS observations that
have been obtained over the lifetime of HST, and the stability inherent to space based observations make this data set ideal for the search of serendipitous stellar
occultations by small objects in the outer solar system.

\begin{figure} [htp]
\centerline{\epsfig{file=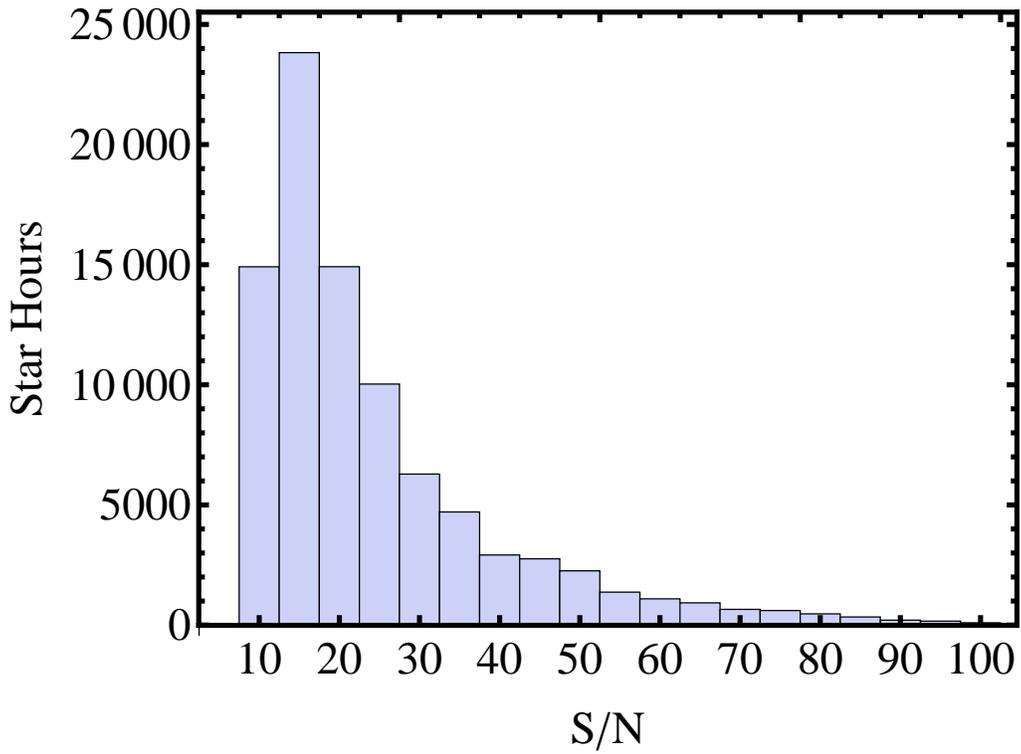, scale=1.2}}
\caption{Distribution of star hours as a function of the mean signal-to-noise
  ratio, $S/N$, in a 40~Hz interval for the entire FGS data set. The signal-to-noise
ratio is defined here as the square root of the mean number of photon counts in a 40 Hz interval.}
\label{fig3}
\end{figure}

\begin{figure} [htp]
\centerline{\epsfig{file=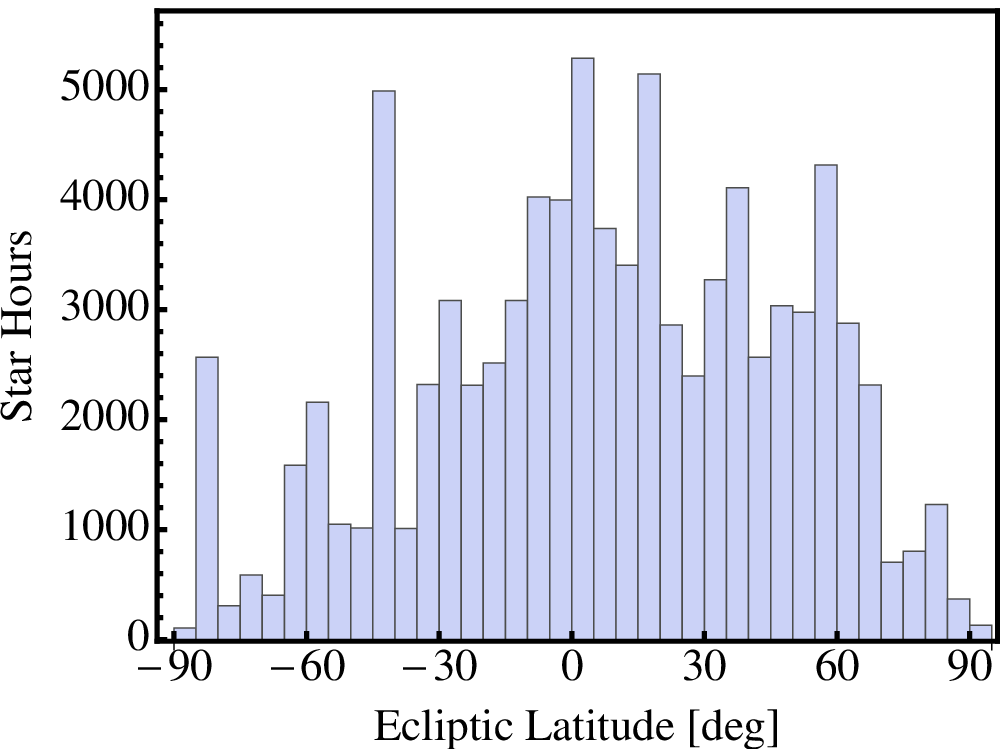, scale=1.2}}
\caption{Distribution of star hours as a function of ecliptic latitude for the
FGS data set.}
\label{fig2}
\end{figure}

\begin{figure} [htp]
\centerline{\epsfig{file=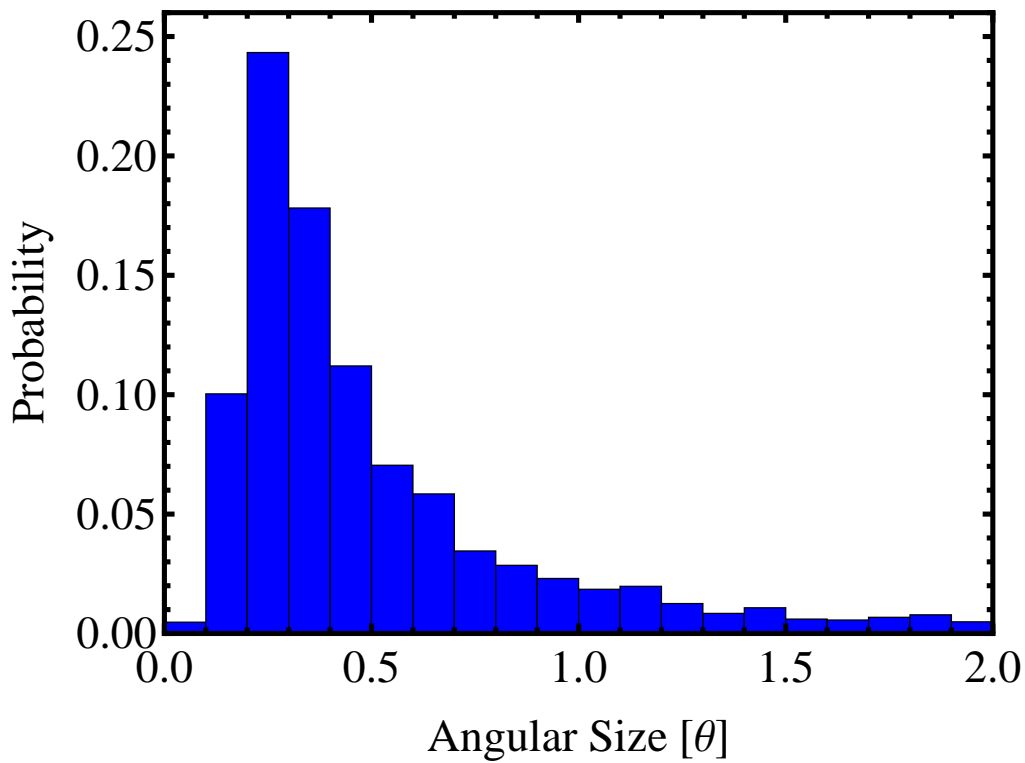, scale=1.2}}
\caption{Distribution of angular sizes of the HST guide stars in our data set,
in units of the angular Fresnel scale. The angular sizes of the guide stars were
calculated by fitting 2MASS JHK and USNO-B1 BR photometry with a black-body
spectrum.}
\label{fig1}
\end{figure}

The entire data set, including low and high ecliptic latitude observations and the data analyzed in \citet{SO09}, consists of $\sim 90,000$ star hours. The signal-to-noise
ratio, defined here as the square root of the mean number of photon counts of all 4 PMTs combined in a 40 Hz interval, as a function of star hours is shown in Figure
\ref{fig3} and ranges from about $10$ to $100$, reflecting the $9 < V<14$ range in magnitude of the HST guide stars. Nominal HST operation uses two
FGSs for guiding, with each FGS observing its own guide star. We are therefore
able to remove false positives due to instrumental effects and, for example,
day/night-time variations due to HST's orbit around the Earth. Figure
\ref{fig2} shows the HST integration time as a function of ecliptic
latitude. About 35\% of the observations are taken in the low ecliptic
latitude region (i.e., $ -20~\rm{\deg} < i < +20~\rm{\deg}$) of the Kuiper
Belt. The remaining high-ecliptic latitude observations provide an excellent
control sample which we use to test for possible false positives.
Figure \ref{fig1} shows the distribution of angular sizes of the HST guide
stars in our data set in units of the Fresnel angular scale at $40~\rm{AU}$ and for
observations at a wavelength of $600~\rm{nm}$. The Fresnel angular scale is given by $\sim \sqrt{\lambda/2a}$, where $\lambda$ is the wavelength of the observations and $a$ the semi-major axis of the Kuiper belt. About 63~\% of the stars in our
data set subtend angular sizes less than 0.5 Fresnel angular scales at a distance of
40~AU, and 84~\% of all stars in our data set have angular sizes of less than
one Fresnel scale. The diffraction pattern that is produced by a sub-km sized
KBO occulting an extended background star is smoothed over the finite stellar
disk. This effect becomes clearly noticeable for stars that subtend sizes
larger than about 0.5 Fresnel scales and reduces the
detectability of occultation events around such stars. The distribution of the
finite angular sizes of the stars are taken into account when we calculate the
detection efficiency of our survey in section \ref{s44}.

Finally, we analyzed the entire data set for correlated noise by calculating the autocorrelation function with lags of 0.025 seconds. Most of the data sets, each consisting of the observations obtained over one HST orbit, are free of statistical significant correlated noise. About 17\% of the roughly 120,000 the data sets contained correlated noise, which is significant at, or above, the $4.5\sigma$ level for a given data set and significant at, or above, the $1\sigma$ level for the analysis of the entire data, which consisted of 120,000 independent trials. This correlated noise is often due to slopes, i.e. long-term variability, in the data. Such long-term trends may affect the results from a bootstrap analysis, however, we show that the two data files containing candidate events for which we use a bootstrap analysis to estimate their significance do not have any statistical significant correlated noise (see section 5 for details).

\section{DATA ANALYSIS}\label{s33}
The 40 Hz time resolution of our survey allows for the detection of the actual
diffraction pattern rather than a simple decrease in the photon counts.  Our
detection algorithm therefore employs a template search using theoretical
light curves and performs $\chi^2$ fitting of the templates to the
data. This template fitting procedure improves the sensitivity of the survey
compared to algorithms that search only for dips in the light curve and aids
with the identification of false positives.

In general, the occultation diffraction pattern is determined by the size of
the KBO, the angular size of the star, the wavelength range of the
observations and the impact parameter between the star and the KBO. 
The typical KBO sizes that our survey is most likely to detect is determined by the signal-to-noise of our data, because small KBOs are more numerous than their larger counterparts and the amplitude of the occultation signal scales as the cross-section of the KBO. Given a typical signal-to-noise ratio of 15 in a 40~Hz interval (see Figure \ref{fig3}), our survey is most likely to detect KBO occultation events caused by objects that are a few hundred meters in radius. Since these objects are significantly smaller than the Fresnel scale, they give
rise to occultation events that are in the Fraunhofer  diffraction regime. This implies
that the diffraction pattern itself is not sensitive to the exact shape of the
KBO, but the amplitude of the diffraction pattern scales linearly as the
cross section of the occulter. This significantly reduces the number of
templates that need to be implemented in the search algorithm. Furthermore, the
templates that we use in our search algorithm treat the stars as point
sources. The reason for this is twofold: First, a large fraction ($\sim
63\rm{\%}$) of the stars in our survey have angular sizes below 0.5 Fresnel
scales, which implies that the point source template is a good approximation
for these stars. Second, using templates with finite angular size stars hurts rather than helps the detection efficiency because the typical uncertainties in the estimated angular radii are too large (i.e. they are typically about 40\%). Since the
diffraction pattern is wavelength dependent, we integrate the light curve
templates that we use in our detection algorithm over a wavelength range of
the FGS observations which extends from 400 to 700~nm. Finally, our search
algorithm includes light curve templates for impact parameters between the KBO
and background star ranging from 0 to 1.0 Fresnel scales in steps of 0.2
Fresnel scales.

For a given impact parameter, our theoretical light curves have three free
parameters. The first is the mean number of photon counts, $m$, which is the
normalization of the light curve. The second is the amplitude of the
occultation, $A$, which is proportional to the cross section of the
KBO, and the third is the duration of the occultation, which is inversely proportional to the relative velocities between HST and the KBO. We
obtain values for the first two parameters from our data by minimizing
$\chi^2$ where
\begin{equation}
\chi^2=\Sigma_{i=1}^{n}\frac{(d_i-\rm{model}_i)^2}{d_i}
\end{equation}
and 
\begin{equation}
\rm{model}_i=m (A(l_i-1)+1)
\end{equation}
where $l_i$ is our theoretical light curve template. We can simply
solve for the values of $A$ and $m$ from our data and therefore only need to
perform a full search the third free parameter, which is the duration of the
occultation and time. The duration of the occultation is independent of the object size, and mainly determined
by the ratio of the Fresnel scale to the relative speed between HST and the
KBO perpendicular to the line of sight. This relative speed is given by
the combination of HST's velocity around the Earth, Earth's velocity around
the Sun and the velocity of the KBO itself. We use this information to
restrict the parameter space for the template widths in our search such that
we are sensitive to KBOs located at the distance of the Kuiper belt between
30\,AU and 60\,AU, but allow for KBO random velocities of up to $v_{\rm{Kepler}}=5$\,km/s.

Our search algorithm identifies occultation candidates by calculating their $\Delta \chi^2$, which we define as the difference between the $\chi^2$ value calculated for a horizontal line, corresponding to no event, and the $\chi^2$ value derived from best fit light curve template. Occultation events have large $\Delta \chi^2$ since they are poorly fitted by a constant straight line, but well matched by the light curve template. If the noise properties were identical over the entire data set, then the probability that a given occultation
candidate is due to random noise can be characterized by a single value of
$\Delta \chi^2$ for all observations. In reality however, the noise
properties are different from observation to observation; especially
non-Poisson tails in the photon counts distribution will give rise to
slightly different $\Delta \chi^2$ distributions. Therefore, ideally, we would
determine a unique detection criterion for each individual HST orbit. However,
this would require to simulate each data set, which contains about an hour of
observations in a single HST orbit, over the entire length of our survey, which is not feasible due to the enormous computational resources that would be required. Instead,
we perform bootstrap simulations over all the FGS data sets together and use this to estimate the typical $\Delta \chi^2$ value that we use in our detection algorithm \citep{E82}. Using bootstrap simulations for estimating the significance of candidate events is justified as long as there is no correlated noise in the data. For
all occultation candidates that exceed this detection threshold, we determined
their statistical significance, i.e. the probability that they are due to
random noise, by extensive bootstrap simulations of the individual data sets. Our detection algorithm flagged all events for which the template fit of the diffraction pattern was better than 15 $\sigma$ and that had a $\Delta \chi^2 >63$. This detection criterion corresponds to about 0.5 false-positive detections over the 19,500 new star hours of low ecliptic latitude observations that are analyzed in this paper. All flagged events that were solely due to one single low or high point were ignored in the data analysis, but included as false-positives when calculating the significance of candidate occultation events. This eliminates a large number of otherwise flagged events that may, for example, be caused by cosmic rays.

\section{DETECTION EFFICENCY}\label{s44}
\begin{figure} [htp]
\centerline{\epsfig{file=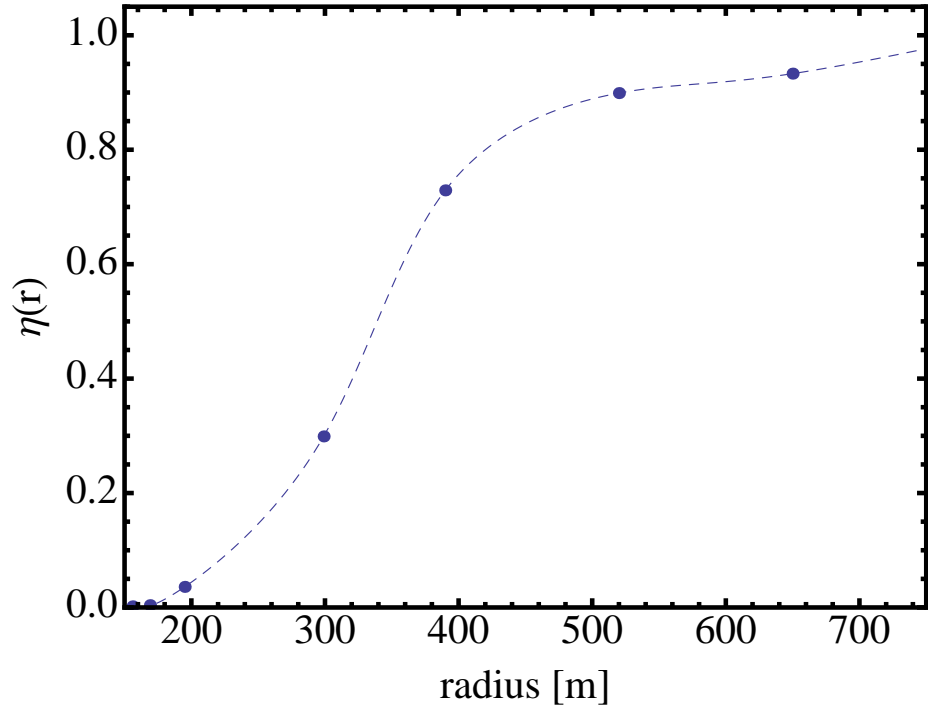, scale=1.2}}
\caption{Detection efficency, $\eta(r)$, as a function of KBO radius. The
  detection efficiency is normalized by an impact parameter equal to 1 Fresnel
  scale, i.e., a detection efficiency of 1 means that the effective
  cross-section for detection has a radius equal to one Fresnel scale. The
  detection efficiency of our survey is $\sim 4$\% for objects with
  $r=200~\rm{m}$ and $75$\% for KBOs with $r=400~\rm{m}$.}
\label{fig4}
\end{figure}

The ability to detect an occultation event of a given size KBO depends on the
impact parameter of the KBO, the duration of the event, the angular size of
the star and the signal-to-noise ratio of the data.  We calculate the
detection efficiency of our survey by planting synthetic events with different radii into the FGS
data and analyzing this modified data set with the same search algorithm that
we used to analyze the original FGS data with the same significance threshold
of $\Delta \chi^2 > 63$. The synthetic events correspond to KBO sizes
ranging from $200\,\rm{m} < r < 850\,\rm{m}$, they have impact parameters from
0 to 2.5 Fresnel scales and a relative velocity distribution that is identical
to that of the actual FGS observations. To account for the finite angular
sizes of the stars we generated light curve templates with stellar angular
radii of 0.1, 0.3, 0.4, 0.6, 0.8 and 1 Fresnel scales. The detection efficiency
of our survey was calculated using the angular size distribution of the FGS
guide stars as shown in Figure \ref{fig1} and the synthetic events were implanted in a subset of the FGS data that had the same signal-to-noise properties as the whole data set shown in Figure \ref{fig3}. Figure \ref{fig4} shows the
detection efficiency, $\eta(r)$, as a function of KBO radius. We normalize our
detection efficiency for a given size KBO to 1 for an effective detection cross
section with a radius of one Fresnel scale. The detection efficiency of our
survey is $\sim 0.04$ for objects with $r=200\,\rm{m}$ and $\sim 0.75$ for KBOs
with $r=400\,\rm{m}$ located at 40\,AU.

\section{RESULTS}\label{s55}

\subsection{One New Occultation Candidate}
\begin{figure} [htp]
\centerline{\epsfig{file=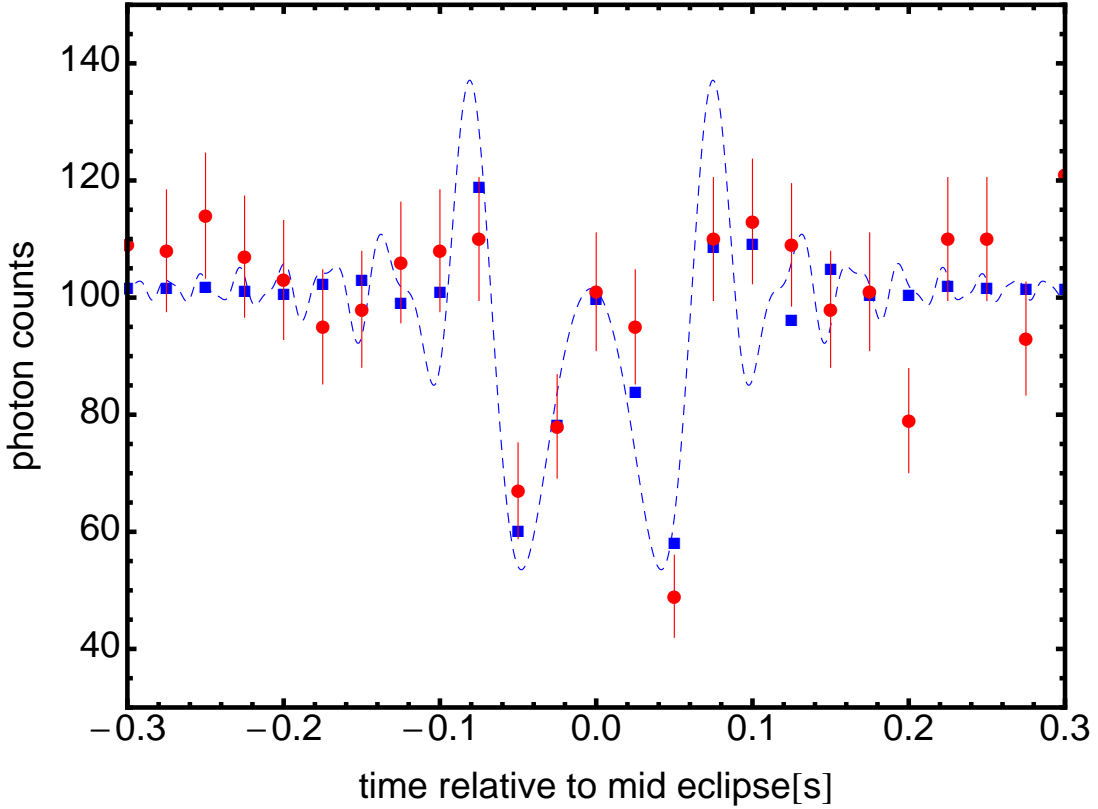, scale=1.2}}
\caption{Photon counts as function of time of the candidate occultation event observed by FGS3. The red points and error bars are the FGS data points with Poisson error bars, the dashed blue line is the theoretical light curve, and the blue squares correspond to the theoretical light curve template in our detection algorithm integrated over 40Hz intervals. We note here that the actual noise for this observation is about 8\% larger than poison noise, due to additional noise sources such as, for example, dark counts, which contribute about 3 to 6 counts per PMT in a 40Hz Interval. The best fit $\chi^2/\rm{dof}$ from our detection algorithm is 27.3/28. The star has an ecliptic latitude of $+6.6^{\circ}$ and its angular radius and effective temperature are $0.58\pm{0.06}$ Fresnel scales and $\sim  5000~{\rm{K}}$, respectively. The position of the star is R.A.=64.74065$^{\circ}$, DEC=28.13064$^{\circ}$ (J2000) and its estimated V-magnitude is 13.9.  Assuming a circular orbit, the best fit parameters yield a KBO size of $r=530\pm{70}$m and a distance of $35\pm{9}$AU.}
\label{fig9}
\end{figure}

Among the $\sim 40$ candidate events that were flagged with a $\Delta \chi^2 > 63$ in the new data set, all but one turned out to be false-positives (see subsection \ref{susfp} for a detailed discussion of the false-positives),
leaving us with one new occultation candidate event. Figure \ref{fig9} shows the candidate event with the best fit template from our search algorithm. The red points and and error bars represent the FGS data with Poisson error bars. The actual noise in this observation is about 8\% larger than Poisson noise, which is due to additional noise sources such as dark counts, which contribute about 3 to 6 counts for a given PMT in a 40Hz interval. The mean signal-to-noise ratio in a 40 Hz interval for this HST orbit of observations is $\sim 10$. The best fit $\chi^2/\rm{dof}$ from our detection algorithm is 27.3/28. Each FGS provides two independent PMT readings and we confirmed that the occultation signature is present in both of the these independent photon counts. The position of the star is R.A.=64.74065$^{\circ}$, DEC=28.13064$^{\circ}$ (J2000), which translates to an ecliptic latitude of $+6.6^{\circ}$. We obtained a spectrum of this star on 12 February 2012 with the Echelle
Spectrograph and Imager (ESI) on the Keck II telescope \citep{Sheines2002}.  We used ESI's echellete mode with a 0.5" slit width, providing spectral coverage from 3900 to 11000 Angstroms
simultaneously with a resolving power $(\lambda/\Delta\lambda)$ of 8100.  We exposed for 300 seconds, achieving a median signal-to-noise of 75 between 6000 and 9000 Angstroms.  Analysis of the stellar spectrum and fitting the $JHK$ bands from 2MASS \citep{SCS06} yields an stellar angular radius and effective temperature of $0.58\pm{0.06}$ Fresnel scales and $\approx  5000~{\rm{K}}$, respectively. Using our best estimate for the stellar angular radius, we find that the best fit parameters yield a KBO size of $r=530\pm{70}$m and, assuming a circular orbit, a distance of $35\pm{9}$AU. We note here, that within the uncertainties of the actual stellar angular radius, that light curve templates with smaller angular radii give a somewhat better fit to the data than templates with larger stellar radii. For objects an circular orbits around the sun, two solutions can fit the duration of the event. However, the second solution corresponds to a distance of 0.2\,AU from the Earth and an objects size of $\approx 50$\,m, and is therefore unlikely. It is also unlikely that the occulting object was located in the Asteroid belt, since the expected occultation rate from Asteroids is about two orders of magnitude less than our implied rate. In addition, an Asteroid would have to have an eccentricity of 0.2 or greater to match the duration of the observed occultation candidate.

\begin{figure} [htp]
\centerline{\epsfig{file=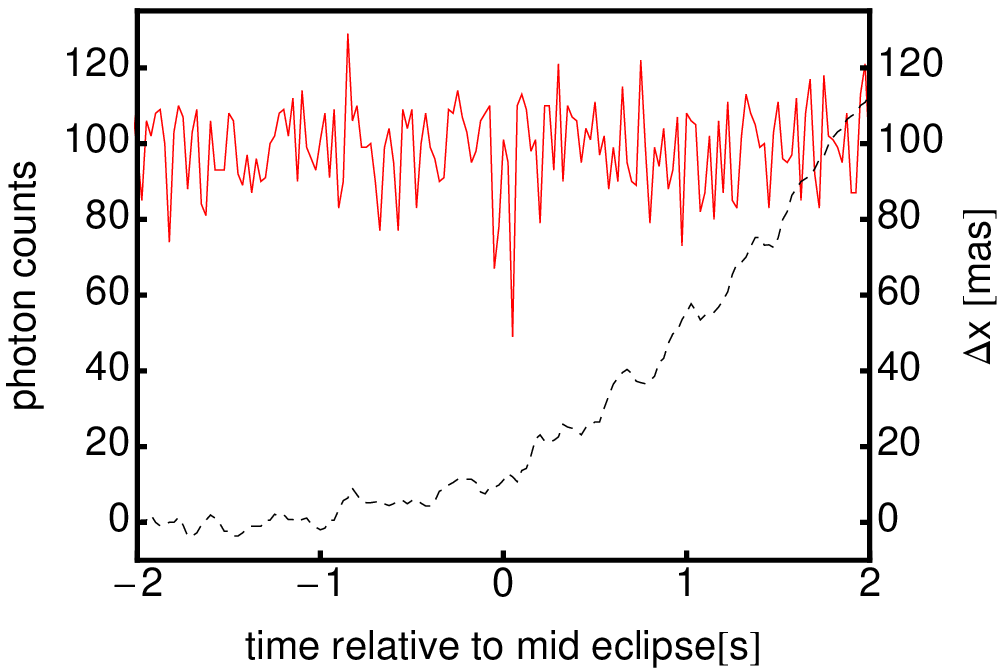, scale=1.2}}
\caption{Photon counts as a function of time during the occultation event observed by FGS3. The red-solid line represents the 40 Hz photon counts from FGS3, while the black-dashed line shows the displacement of the guide star along the x-axis, $\Delta x$, in the FGS relative to its position at time = -2 seconds. The occultation event occurred (at time=0) approximately one second after HST began a commanded small angle maneuver associated with a planned displacement of the active science instrument's aperture on the sky. During this maneuver FGS3 actively tracked its guide star, keeping it at interferometric null, just as it did prior to the maneuver. Thus the decrease of photon counts near time = 0 is not correlated with or caused by the telescope repointing. Note, the displacement along the FGS y-axis was far smaller and is therefore not plotted here.}
\label{fig10}
\end{figure}

\begin{figure}[htp]
\centerline{\epsfig{file=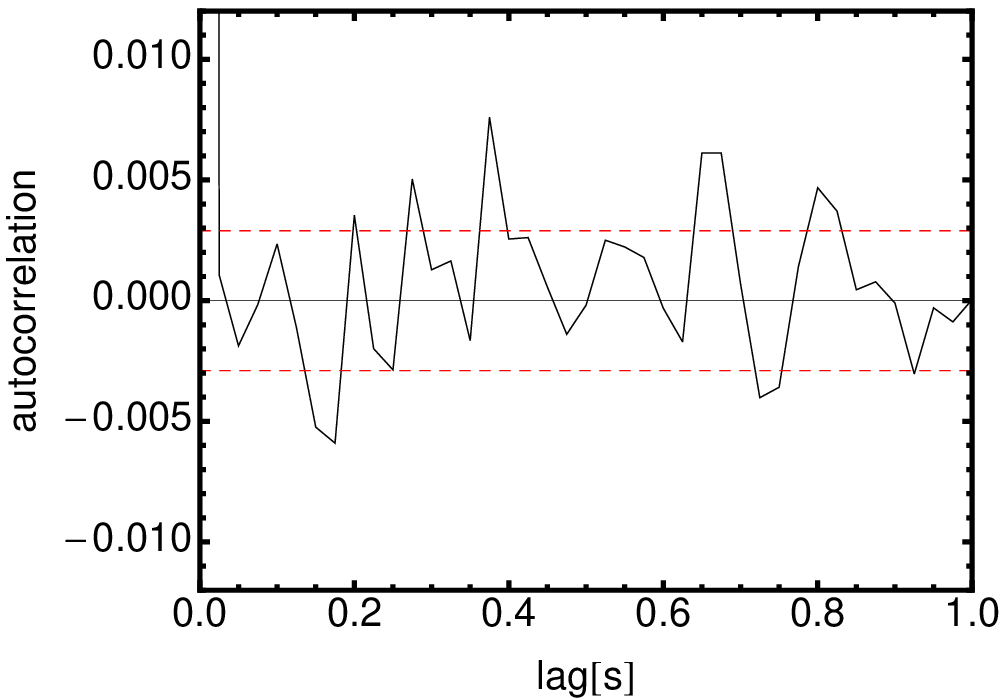, scale=1.3}}
\caption{Autocorrelation function for lags between 0 and 1 seconds for the HST orbit of observations containing the candidate event shown in Figure \ref{fig9}. We note that the autocorrelation at lag 0 is, by definition, equal to one. The red-dotted lines shows the upper and lower $1\sigma$ errors of the autocorrelation. The autocorrelation function of this data is consistent with zero.}
\label{fig133}
\end{figure}

\begin{figure} [htp]
\centerline{\epsfig{file=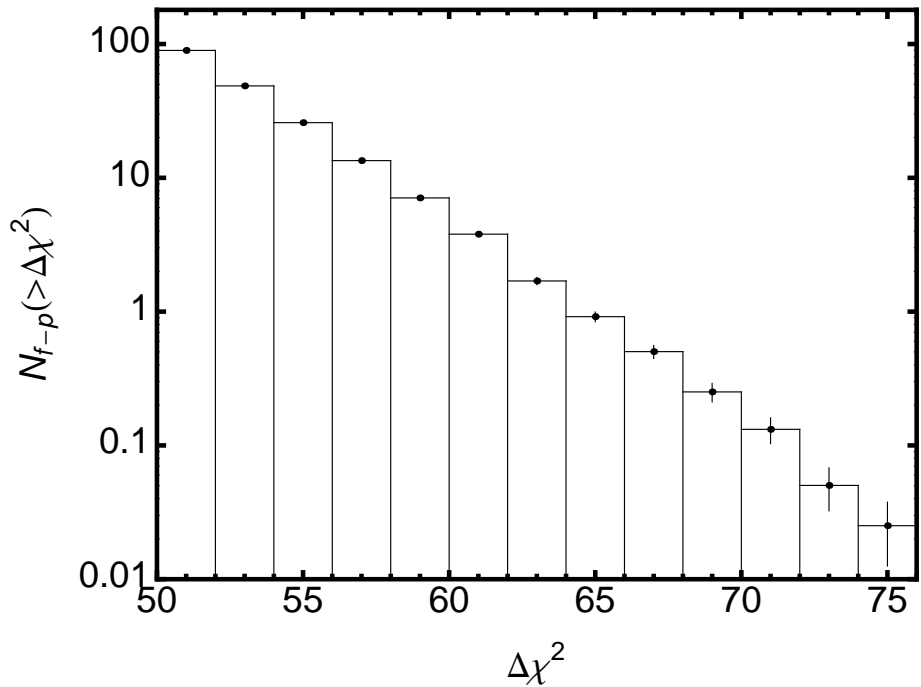, scale=1.2}}
\caption{Cumulative number of false-positives, $N_{f-p}$ as a function of $\Delta \chi^2$ for the candidate event normalized to 19,500 star hours, which correspond to the length of the entire low ecliptic latitude observations analyzed in this paper. The false-positives were obtained from bootstrap simulations using the one orbit of HST observations ($\sim$ 50 minutes) in which we found the candidate event. From these bootstrap simulations we find a probability of $\approx 5 \%$ that events like the occultation candidate with $\Delta \chi^2 = 71.9$ are caused by random statistical fluctuations in the entire low ecliptic latitude data set.}
\label{fig14}
\end{figure}

We estimate the probability that this candidate event is due to statistical fluctuations using bootstrap simulations.
This approach is justified as long as there is no correlated noise in the data. We calculate the autocorrelation for lags between 0 and 1 seconds for the HST orbit of observations that contained the event and find that the autocorrelation function is consistent with zero (see Figure \ref{fig133}). Using the data from the HST orbit, that contained the event, we removed the event itself and simulated $3.1 \times 10^6$ star hours, which corresponds to 161 times the low ecliptic latitude observations analyzed in this paper. This calculation required $\sim 2300$ CPU days of computing power. Figure \ref{fig14} shows the cumulative number of false-positives, $N_{f-p}$ as a function of $\Delta \chi^2$. The number of false-positives was normalized to 19,500 star hours, which correspond to the length of the entire low ecliptic latitude observations analyzed here. In the entire bootstrap analysis we obtained 8 events with a $\Delta \chi^2 \geq 71.9$. This implies a probability of  $\sim 5 \%$ that events like the occultation candidate with $\Delta \chi^2 = 71.9$ are caused by random statistical fluctuations over the entire low ecliptic latitude observations analyzed in this paper.

\subsection{False-Positives}\label{susfp}
\begin{figure} [htp]
\centerline{\epsfig{file=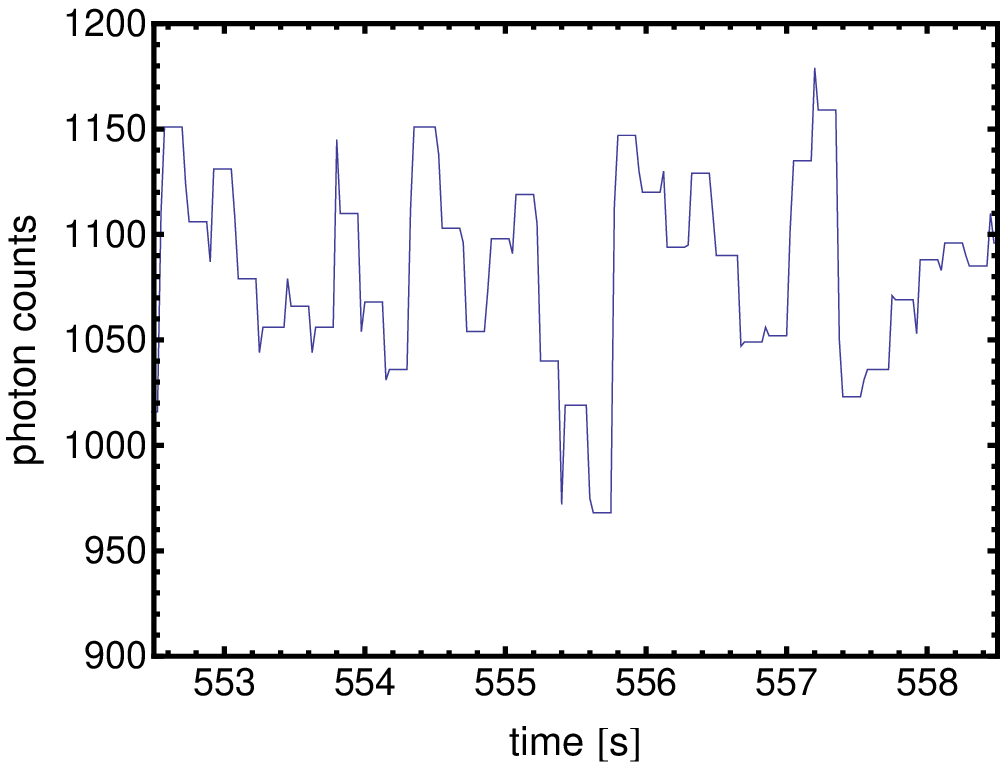, scale=1.1}}
\caption{Photon counts as a function of time. An example of a false-positive found in the FGS data, which is due to what looks like a slower read out of the photon counts. False-positives like this one are due to update problems that are encountered as the telemetry from HST goes through various transfer stations and ground stations on its way to the archive. }
\label{fig8}
\end{figure}

\begin{figure} [htp]
\centerline{\epsfig{file=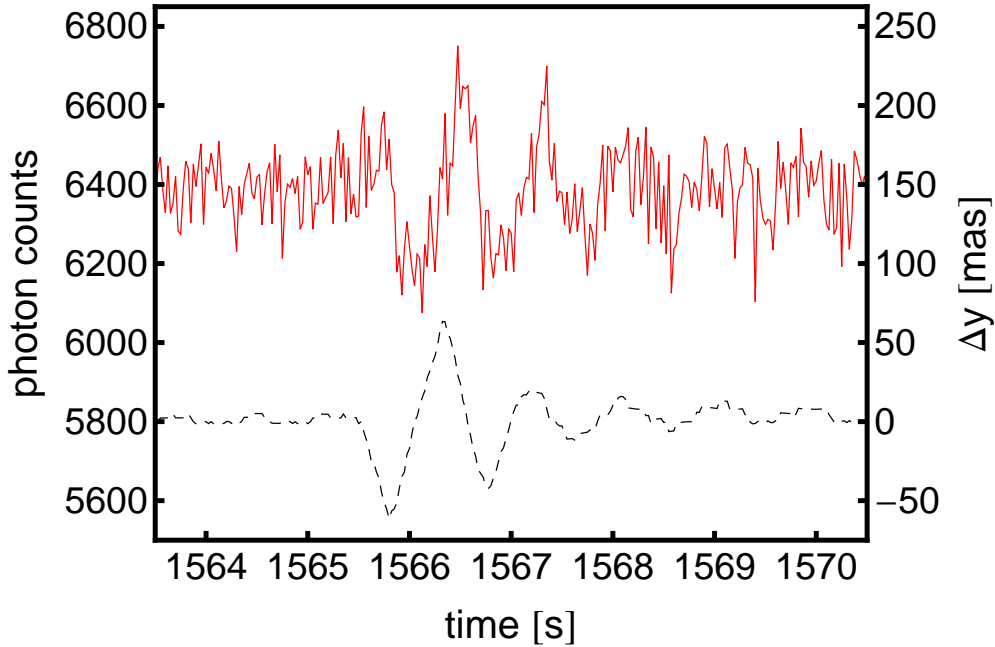, scale=1.3}}
\caption{An example of a false-positive that is due to jitter introduced by the displacement of the guide star from its null position. The red-solid line represents the photon counts of the FGS data and the black-dashed line the corresponding displacement of the guide star relative to its null position along the y-axis, as recorded by the FGS, as a function of time.}
\label{fig6}
\end{figure}

Among the $\sim 40$ candidate events that were flagged because they had a $\Delta \chi^2 > 63$, all but one turned out to be false-positives. The most common false-positives were due to what looks like a slower read out of the photon counts (see Figure \ref{fig8}) or showed a correlation between the signature in the photon counts and the displacement of the guide star from its null position on the FGS (see Figure \ref{fig6}).
Figure \ref{fig6} shows an example of a false-positive that shows a strong correlation between the number of photon counts and the displacement of the guide star from its null position on the FGS. The jitter introduced due to the displacement of the guide star from its null position causes an up to $3\%$ change in the photon counts and is therefore only detected by our search algorithm for stars that have photon counts above a few thousand in a single 40Hz interval. The jitter due to the guide star's displacement has a characteristic frequency of 1Hz and is therefore easy to identify.

\subsection{High-Ecliptic latitude Control Sample}\label{hec}
\begin{figure} [htp]
\centerline{\epsfig{file=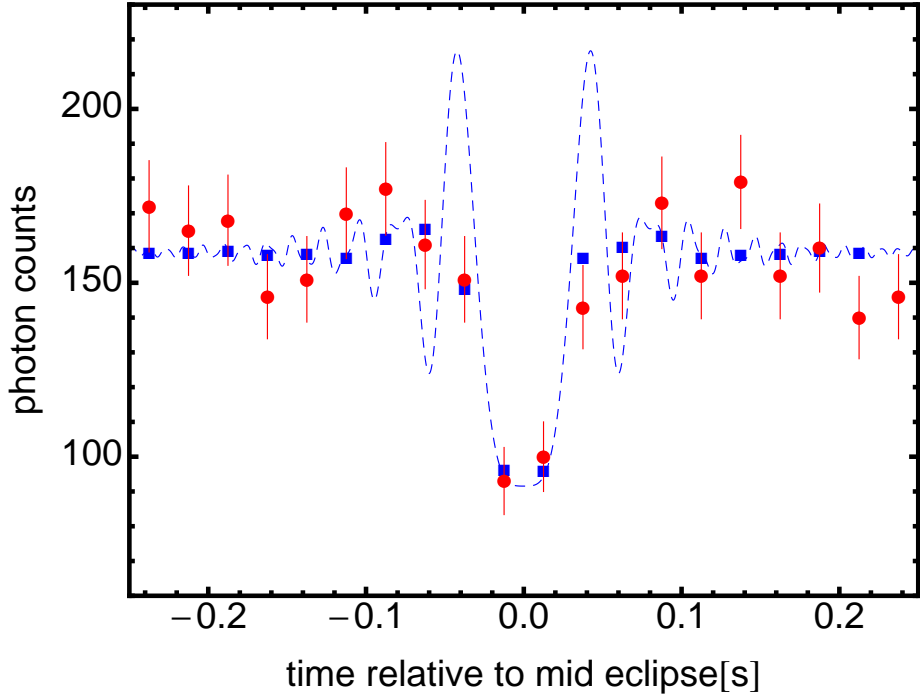, scale=1.2}}
\caption{Photon counts as function of time of the only flagged event in our control sample that could not be contributed to either jitter, induced due to the displacement of the guide star from its null position (see Figure \ref{fig6}), or a lower read out frequency of the PMTs (see Figure \ref{fig8}). The red points and error bars are the FGS data points with Poisson error bars, the dashed blue line is the theoretical light curve, and the blue squares correspond to the theoretical light curve integrated over 40Hz intervals. The best fit $\chi^2/\rm{dof}$ is 13.5/17. Bootstrap simulations yield a probability of $\sim 21 \%$ that events like this with $\Delta \chi^2 = 72.3$ are caused by random statistical fluctuations over the high ecliptic latitude control sample analyzed in this paper. This implies a $\sim 79 \%$ chance that this flagged event is due to a high inclination KBO, but this interpretation seems unlikely given the known inclination distribution of large KBOs.}
\label{fig22}
\end{figure}

\begin{figure} [htp]
\centerline{\epsfig{file=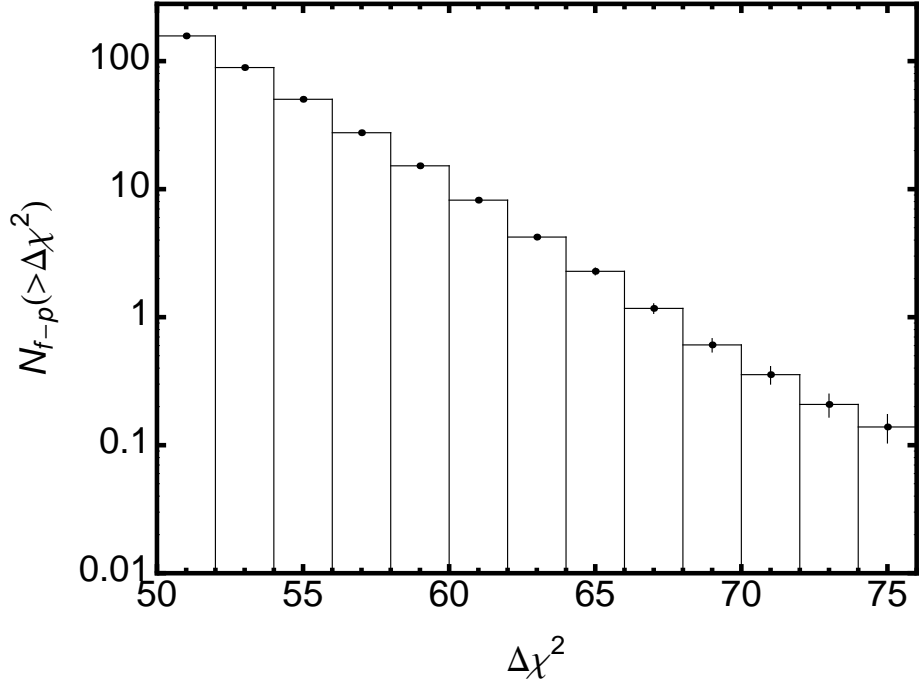, scale=1.2}}
\caption{Cumulative number of false-positives, $N_{f-p}$ as a function of $\Delta \chi^2$ for the flagged event found in the high ecliptic latitude control sample. The number of false-positives is normalized to 36,000 star hours, which correspond to the length of the entire high ecliptic latitude observations analyzed in this paper. The false-positives plotted here were obtained from bootstrap simulations using the one orbit of HST observations in which this event occurred. From these bootstrap simulations we find a probability of $\sim 21 \%$ that events like this with $\Delta \chi^2 = 72.3$ are caused by random statistical fluctuations over the high ecliptic latitude control sample analyzed in this paper.}
\label{fig21}
\end{figure}

In addition to the 19,500 star hours of low ecliptic latitude observations, $|b| < 20^{\circ}$, we also analyzed $\sim 36,000$ star hours of high ecliptic latitude, $|b| > 20^{\circ}$, observations and used these as a control sample. We analyzed the high ecliptic latitude observations with exactly the same detection algorithm as the low ecliptic latitude data. A total of $\sim 70$ candidate events were flagged with $\Delta \chi^2 > 63$. All but one of these events were due to either jitter, induced due to the displacement of the guide star from its null position (see Figure \ref{fig6}), or a lower read out frequency of the PMTs (see Figure \ref{fig8}). The only event which was not caused by either of these two effects is shown in Figure \ref{fig22}. The ecliptic latitude of this event is $81.5^{\circ}$ and it has a $\Delta \chi^2=72.3$. Calculation of the autocorrelation function of this orbit of HST observations showed that there is no statistical significant correlated noise in the data. Using bootstrap simulations we find a $21\%$ probability that this events is due to random statistical fluctuations (see Figure \ref{fig21}), given the entire high ecliptic latitude data set analyzed in this paper. This implies a $\sim 79 \%$ chance that this flagged event is due to a high inclination KBO, but this interpretation seems unlikely given the known inclination distribution of large KBOs.

\section{Discussion \& Conclusions}\label{s66}
We combine the one candidate KBO occultation event presented in this paper, which we found in 19,500 star hours of low ecliptic latitude HST-FGS observations, with the single event that was reported by \citet{SO09}. \citet{SO09} analyzed 12,000 star hours of low ecliptic latitude HST-FGS observations and reported one KBO occultation event at an ecliptic latitude of +14.4$^{\circ}$. First we test whether the ecliptic latitudes of the two events are consistent with the observed inclination distribution of larger KBOs and then we use the two events to estimate the abundance of sub-km sized KBOs.

\subsection{Inclination Distribution}
In this subsection, we test if the observed ecliptic latitude distribution of the two occultation events is consistent with the inclination distribution of larger KBOs inferred from direct searches.

\begin{figure} [htp]
\centerline{\epsfig{file=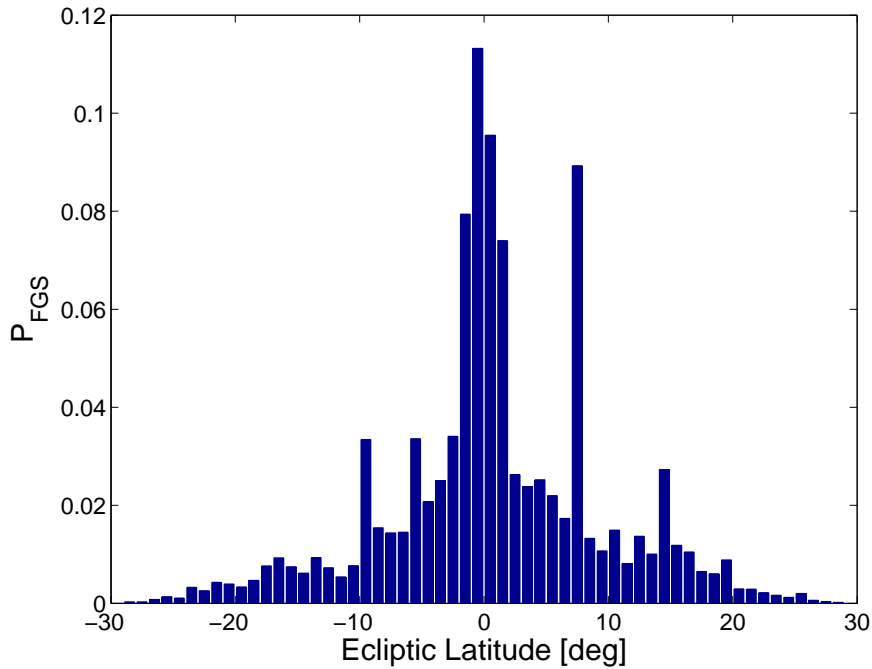, scale=0.63}}
\caption{Occultaion detection probability per degree, $P_{FGS}$, as a function of ecliptic latitude for the low ecliptic latitude observations $|b| < 20^{\circ}$ of the FGS data set. The detection probability was calculated from the ecliptic latitude distribution of the FGS data shown in Figure \ref{fig2}. Note, we assumed that the KBO ecliptic latitude distribution is symmetric about the ecliptic and ignored the small $1.6^{\rm{\circ}}$ inclination of the Kuiper belt plane relative to the ecliptic.}
\label{fig12}
\end{figure}

We first calculate for each star in our data set the ecliptic latitude, and the amount of time it was observed by the FGSs. Figure \ref{fig2} shows the distribution of star hours as a function of ecliptic latitude. We denote this function by $t_{{\rm FGS}}(\beta)$. Using the inclination distribution of KBOs, $P(i_{{\rm KBO}})$, from  \citet{E05}, we randomly draw inclinations from $P(i_{{\rm KBO}})$ and for each random declination we choose a random ecliptic latitude $\sin(\beta)=\sin(i)\sin(\lambda)$ \footnote{Specifically, we use the numerical inclination distribution presented in Fig 20. of \citet{E05}}. Here the ecliptic longitude, $\lambda$, is a random number drawn from a uniform distribution between 0 and $2\pi$. This give us the probability distribution of the instantaneous ecliptic latitude distribution, $P_{{\rm KBO}}(\beta)$, associated with inclination distribution from \citet{E05}. Figure \ref{fig12} shows the product of $P_{{\rm KBO}}(\beta)$ and $t_{{\rm FGS}}(\beta)$ (denoted by $P_{{\rm FGS}}(\beta)$), and gives the probability that our survey will detect a KBO at a given ecliptic latitude, assuming the KBOs are drawn from the \citet{E05} inclination distribution. We assumed here that KBOs follow circular orbits and that the S/N distribution of the FGS data is the same at all ecliptic latitudes.

\begin{figure} [htp]
\centerline{\epsfig{file=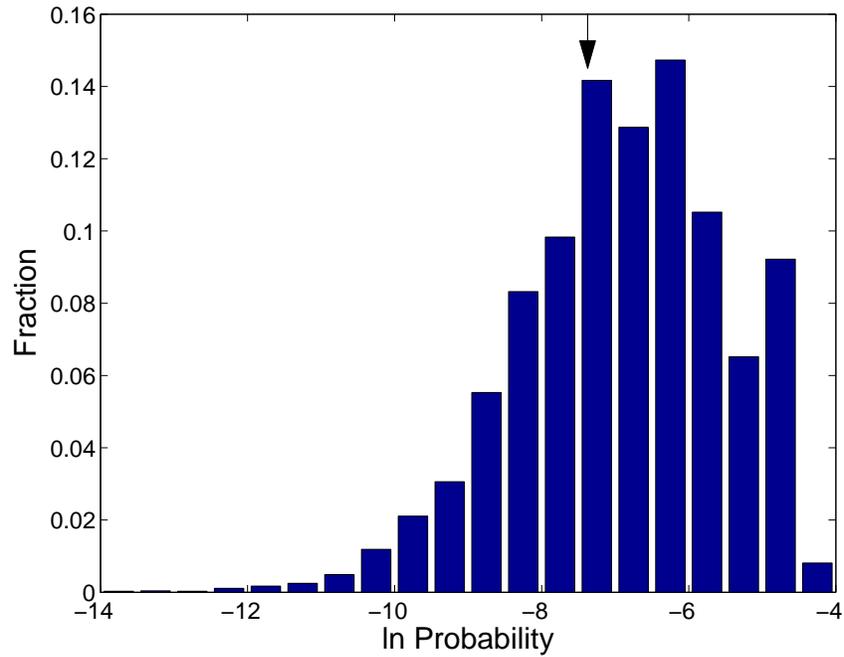, scale=0.63}}
\caption{Probability distribution derived from Monte-Carlo simulations, which consisted of drawing 10,000 randoms pairs of of ecliptic latitudes from the distribution shown in Figure \ref{fig12} and calculating the probability for each pair. The arrow denotes the log-probability value of the  of our observations, suggesting that our observed ecliptic latitudes are consistent with ecliptic latitudes drawn from the inclination distribution of large KBOs.}
\label{fig13}
\end{figure}

In order to see if our observations are consistent with the inclination distribution of large KBOs, we calculate the probability distribution for observing two events at different latitudes using the detection probability as a function of ecliptic latitude shown in Figure \ref{fig12}. The log-probability of our ecliptic latitude observations is: $\sum_{j}^{2}{ \ln{P_{\rm FGS}(\beta_{\rm j})}}$, where $\beta_{\rm j}$ are our observed ecliptic latitudes of $+6.6^{\rm{\circ}}$ and $+14.4^{\rm{\circ}}$. In order to put this into context we need to estimate the log-probability distribution given our data set. This is done using Monte-Carlo simulations, which consist of drawing 10,000 randoms pairs of ecliptic latitudes and calculating the log-probability for each pair. Figure~\ref{fig13} shows the probability distribution, where the arrow denotes the value of the log-probability of our observations. This plot suggests that our observed ecliptic latitudes are consistent with ecliptic latitudes drawn from the inclination distribution of large KBOs. However, this conclusion would change if the flagged event in our control sample (discussed in section \ref{hec}) was due to an actual KBO.

\subsection{Abundance of Sub-km sized KBOs}
We now combine the results from this paper with the work of \citet{SO09} to estimate the abundance of sub-km sized KBOs.

\begin{figure} [htp]
\centerline{\epsfig{file=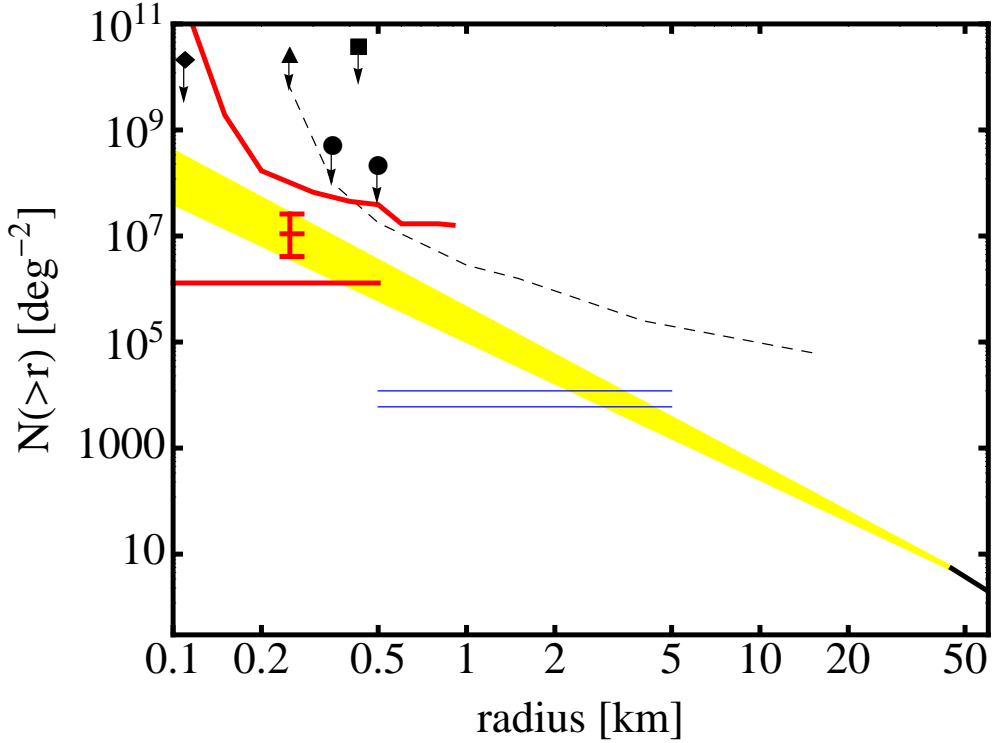, scale=1.3}}
\caption{Cumulative KBO size distribution as a function of KBO radius. The results from our FGS survey are presented in three different ways: (i) the red point plotted at $r$=250\,m with the upper and lower error bars gives the best estimate of the KBO surface density around the ecliptic ($-5^{\circ} < b < 5^{\circ}$) from our survey with $1\sigma$ errors, assuming that sub-km sized KBOs follow the same inclination distribution as their larger, 100-km sized, cousins.  (ii) The upper and lower red curves correspond to our upper and lower 95\% confidence level. (iii) The $1\sigma$ range for our best estimate of the power law size distribution index, $q=3.8\pm{0.2}$ is given by the yellow shaded region normalized to $N(>r) = 5.4\,\rm{deg}^{-2}$ at a radius of 45km \citep{FH08}. The plotted red point, the upper and lower limits and the power law size distributions were all derived assuming that sub-km sized KBOs have the same ecliptic latitude distribution as their larger, 100km-sized, counterparts. All would be a factor of 2.4 lower if sub-km sized KBOs would have an ecliptic latitude distribution that is close to uniform for $-20^{\circ} < b < 20^{\circ}$. The area between the two blue horizontal lines defines the required scattered disk KBO abundances from \citet{VM08} in order to supply the Jupiter Family comets. In addition, 95\% upper limits from various optical KBO occultation surveys are shown as black symbols with arrows (circles \citet{BP09}, square \citet{BKW08}, triangle \citet{WP10}, and diamond \citet{RDD06}) and as a dashed back line for TAOS \citep{BZ10}.}
\label{fig20}
\end{figure}

The number of occultation events is given by
\begin{equation}
N_{events} = -2 v_{rel} F \int_{r_{min}}^{r_{max}} \int_{-b}^{b}\eta(r)
  \frac{\Delta t}{\Delta b} \frac{dN(r,b)}{dr} \,\mathrm{d}b
  \mathrm{d}r
\label{eq1}
\end{equation}
where $v_{rel}=22\,\rm{km/s}$ is the typical relative velocity between the KBO
and the observer, $b$ is the ecliptic latitude, $\Delta t/\Delta b$ is the
time observed per degree in ecliptic latitude, as shown in Figure \ref{fig2}, and $F=1.3\rm{\,km}$ is the Fresnel scale. The sky density of KBOs is both a function of ecliptic latitude, $b$, and the KBO radius, $r$. Therefore, in order to estimate the total number of
KBOs of a given size or their corresponding sky density, we need to make an
assumption regarding their ecliptic latitude distribution. Unfortunately
very little is currently know about the ecliptic latitude distribution of sub-km sized
objects in the Kuiper belt. We therefore estimate the total number of KBOs for two very different ecliptic latitude distributions, currently both are consistent with the ecliptic latitudes of the candidate event presented in this paper and the event reported by \citet{SO09}. In the first case we assume that the KBO latitude distribution, $f(b)$, is uniform between $-20^{\circ}$ and $+20^{\circ}$, such that $f(b)=1$ for $-20^{\circ} < b< 20^{\circ}$ and zero otherwise. In the second case we assume that small sub-km sized KBOs
follow the same ecliptic latitude distribution as their larger $
100$~km-sized counterparts and use the ecliptic latitude distribution provided
in \cite{E05} in Figure 14. We further assume that the KBO size distribution follows a
power law, such that it can be written as $N(r,b)=n_0 r^{-q+1}
f(b)$ where $n_0$ is the normalization factor for the cumulative surface
density of KBOs. Substituting for $dN(r,b)/dr$ in Equation \ref{eq1} and
solving for $n_0$ we get
\begin{equation}
n_0= \frac{N_{events}}{2 v_{rel} F (q-1) \int_{r_{min}}^{r_{max}}
  \eta(r) r^{-q} \,\mathrm{d}r \int_{-b}^{b}
  f(b)\frac{\Delta t}{\Delta b} \,\mathrm{d}b}.
\label{eq3}
\end{equation}
Evaluating Equation \ref{eq3} assuming a uniform KBO ecliptic latitude distribution for $-20^{\circ} < b< 20^{\circ}$ yields a cumulative KBO surface density of
\begin{equation}\label{e21}
N(r>250\,m) = 4.4^{+5.8}_{-2.8} \times 10^6\,\rm{deg^{-2}}.
\end{equation}
Similarly, evaluating Equation \ref{eq3}, assuming that the small, sub-km sized KBOs follow the ecliptic latitude distribution from \citet{E05}, yields a cumulative KBO surface density averaged
over the ecliptic ($\vert b \vert < 5^{\circ}$) of
\begin{equation}\label{e22}
N(r>250\,m) = 1.1^{+1.5}_{-0.7} \times 10^7\,\rm{deg^{-2}}.
\end{equation}
When evaluating the integral over $r$ in Equation \ref{eq3}, we assumed $q=4$. We note however that the value for the cumulative KBO surface density at $r$=250\,m only weakly
depends on the exact choice for $q$. We therefore find an ecliptic KBO abundance for bodies with $r>250\,\rm{m}$ that ranges, depending on the actual inclination distribution of sub-km sized KBOs, between $4.4 \times 10^6\,\rm{deg^{-2}}$ to $1.1 \times 10^7\,\rm{deg^{-2}}$. This is the best measurement of the surface density of sub-km sized KBOs to date and about a factor of 2 lower than the first results published by \citet{SO09}. Figure \ref{fig20} displays the results from the FGS survey and summarizes published upper limits from various works. The red point plotted at $r$=250\,m with the upper and lower error bars in Figure \ref{fig20} gives the best estimate of the KBO surface density around the ecliptic ($-5^{\circ} < b < 5^{\circ}$) from our survey with $1\sigma$ errors, assuming that sub-km sized KBOs follow the same inclination distribution as their larger, 100-km sized, cousins. The upper and lower red curves correspond to our upper and lower 95\% confidence level, which are derived without assuming any size distribution. This limit and the red point would be a factor of 2.4 lower if sub-km sized KBOs would have an ecliptic latitude distribution that is close to uniform for $-20^{\circ} < b < 20^{\circ}$. The jump between 500~m and 600~m in the upper limit curve is due to the fact that below 500~m it is calculated for 2 events, whereas above 600~m for no events. The 95\% upper limit from TAOS \citep{BZ10} is about a factor of 2 lower than the one derived from our FGS survey, if sub-km sized KBOs have the same ecliptic latitude distribution as their larger, 100km-sized, counterparts, and about the same if they follow a uniform ecliptic latitude distribution between $-20^{\circ}$ and $+20^{\circ}$.

Assuming that the KBO size distribution can be well described by a single power law that is parameterized by $N(>r) \propto r^{1-q}$, where $N(>r)$ is the number of KBOs with radii greater than $r$, and $q$ is the power law index, we can use the above estimated KBO abundances to calculate the power law index below the observed break in the KBO size distribution. Assuming a break radius of 45km and a corresponding cumulative KBO surface density of $5.4\rm{deg}^{-2}$ \citep{FH08} we find $q=3.6\pm{0.2}$ and $q=3.8\pm{0.2}$ for a uniform KBO ecliptic latitude distribution and a KBO latitude distribution from \citet{E05}, respectively. 

\citet{T10} found that, in contrast to larger KBOs with sizes above the break radius, fainter ($R > 26$) KBOs, which have sizes below the break radius, are dominated by dynamically excited objects ($i>5^{\circ}$). If this result applies all the way to sub-km sized KBOs, then this suggests that the inclination distribution for sub-km sized KBOs may be dominated by dynamically excited objects and that the true abundance of sub-km sized KBOs, and their corresponding size distribution power-law index, may in fact lie somewhere between $N(r>250\,m) = 4.4 \times 10^6 - 1.1 \times 10^7\,\rm{deg^{-2}}$, and $q=3.6-3.8$, that we estimated above (see Figure \ref{fig20}). In addition, we can rule out a  single power law below the break  with $q>4.0$ at 2$\sigma$.

\subsection{Conclusion}
We presented here the analysis of $\sim 19,500$ new star hours of low ecliptic latitude ($|b| \leq 20^{\rm{\circ}}$) archival data that was obtained by the HST over a time span of more than nine years. Our search for stellar occultations by small Kuiper belt objects (KBOs) yielded one new candidate event, which, assuming a circular orbit, corresponds to a body with a $\sim 500\,\rm{m}$ radius located at a distance of about 40~AU. Using bootstrap simulations, we estimate a probability of $\approx 5\%$, that this event is due to random statistical fluctuations within the analyzed data set. Combining this new event  with the single KBO occultation reported by \citet{SO09} we show that their ecliptic latitudes of $6.6^{\circ}$ and $14.4^{\circ}$, respectively, are consistent with the observed inclination distribution of larger, 100-km sized  KBOs. 

Assuming that the new candidate event and the event previously reported by \citet{SO09} are indeed genuine KBO occultations and that small, sub-km sized KBOs have the same ecliptic latitude distribution as larger KBOs, we find an ecliptic surface density of KBOs with radii larger than 250m of $1.1^{+1.5}_{-0.7} \times 10^7\,\rm{deg^{-2}}$. If sub-km sized KBOs have instead a uniform ecliptic latitude distribution for $-20^{\circ} < b< 20^{\circ}$ we find $N(r>250\,m) = 4.4^{+5.8}_{-2.8} \times 10^6\,\rm{deg^{-2}}$. The ecliptic latitudes of the two events are consistent with both a uniform ecliptic latitude distribution and the ecliptic latitude distribution of larger KBOs published by \citet{E05}. These estimated KBO abundances provide the best measurements of the abundance of sub-km sized KBOs to date and are, although consistent within 1$\sigma$, about a factor of 2 lower than previous results published by \citet{SO09}.

Assuming that the KBO size distribution for bodies with radii smaller than the break radius can be well described by a single power law that is parameterized by $N(>r) \propto r^{1-q}$, where $N(>r)$ is the number of KBOs with radii greater than $r$, and $q$ is the power law index, we find $q=3.6\pm{0.2}$ and $q=3.8\pm{0.2}$ for a uniform KBO ecliptic latitude distribution and a KBO ecliptic latitude distribution that follows the observed distribution for large KBOs, respectively. These results are consistent with a power-law index of $q=3.5$, corresponding to a collisional cascade consisting of material strength dominated bodies \citep{D69} that all have the same constant velocity dispersion, within better than 1 and 2$\sigma$, respectively. We caution however that the actual size distribution of small KBOs is likely to exhibit significant deviations from a single power law due to possible changes from gravity to martial strength dominated bodies and waves that may exist in the small KBO size distribution \citep[e.g.][]{OBG03,PS05}. In addition, the velocity dispersion of the bodies in the cascade may not be constant and instead evolve as a function of size, which results in a size distributions that is significantly steeper than the one derived without velocity evolution (e.g. the standard q = 3.5 power-law index of the Dohnanyi differential size spectrum can change to an index as large as q = 4) \citep{PS12}.

Finally, our findings suggest that small KBOs are numerous enough to satisfy the required supply rate for the Jupiter family comets calculated by \citet{VM08} for scattered disk objects. We can rule out a  single power law below the break  with $q>4.0$ at 2$\sigma$ confirming a strong deficit of sub-km sized KBOs compared to a population extrapolated from objects with $r> 45~\rm{km}$. This suggests that small KBOs are undergoing collisional erosion and that the Kuiper belt is a true analogue to the dust producing debris disks observed around other stars.

\acknowledgements{We thank Dr. Evan Kirby for analyzing and fitting the guide star spectrum. For HS support for this work was provided by NASA through
  Hubble Fellowship Grant \# HST-HF-51281.01-A awarded by the Space Telescope
  Science Institute, which is operated by the Association of Universities for
  Research in Astronomy, Inc., for NASA, under contact NAS 5-26555. RS acknowledges support by an ERC grant, a Packard Fellowship and HST Grant \# HST-AR-12154.08-A. EOO is incumbent of the Arye Dissentshik career development chair and is grateful to support by a grant from the Israeli Ministry of Science and to support from
the The Helen Kimmel Center for Planetary Science.}

\end{document}